\documentclass[aps,twocolumn, amsmath,amssymb, superscriptaddress, reprint]{revtex4-1}

\usepackage{hyperref}

\usepackage[utf8]{inputenc}
\usepackage{lmodern}
\usepackage[version=3]{mhchem}
\usepackage{amsmath,amssymb,amstext,amsfonts}
\usepackage{natbib}
\usepackage{psfrag,graphicx}
\usepackage[]{units}
\usepackage{upgreek}
\usepackage{textcomp} 
\usepackage{color}
\usepackage{float}
\usepackage{subfigure}
\usepackage{array} 
\usepackage{epstopdf}
\usepackage[ngerman, num]{isodate}

\begin{document}
\title{Topology Optimized and 3D Printed Polymer Bonded Permanent Magnets for a Predefined External Field}

\author{C. Huber}
\thanks{Correspondence to: \href{mailto:christian.huber@tuwien.ac.at}{christian.huber@tuwien.ac.at}}
\affiliation{Institute of Solid State Physics, Vienna University of Technology, 1040 Vienna, Austria}
\affiliation{Christian Doppler Laboratory for Advanced Magnetic Sensing and Materials, 1040 Vienna, Austria}

\author{C. Abert}
\affiliation{Institute of Solid State Physics, Vienna University of Technology, 1040 Vienna, Austria}
\affiliation{Christian Doppler Laboratory for Advanced Magnetic Sensing and Materials, 1040 Vienna, Austria}

\author{F. Bruckner}
\affiliation{Institute of Solid State Physics, Vienna University of Technology, 1040 Vienna, Austria}
\affiliation{Christian Doppler Laboratory for Advanced Magnetic Sensing and Materials, 1040 Vienna, Austria}

\author{M. Groenefeld}
\affiliation{Magnetfabrik Bonn GmbH, 53119 Bonn, Germany}

\author{I. Teliban}
\affiliation{Magnetfabrik Bonn GmbH, 53119 Bonn, Germany}

\author{C. Vogler}
\affiliation{Institute of Solid State Physics, Vienna University of Technology, 1040 Vienna, Austria}

\author{D. Suess}
\affiliation{Institute of Solid State Physics, Vienna University of Technology, 1040 Vienna, Austria}
\affiliation{Christian Doppler Laboratory for Advanced Magnetic Sensing and Materials, 1040 Vienna, Austria}

\date{\today}

$ $\newline

\begin{abstract}
Topology optimization offers great opportunities to design permanent magnetic systems that have specific external field characteristics. Additive manufacturing of polymer bonded magnets with an end-user 3D printer can be used to manufacture permanent magnets with structures that have been difficult or impossible to manufacture previously. This work combines these two powerful methods to design and manufacture permanent magnetic system with specific properties. The topology optimization framework is simple, fast, and accurate. It can be also used for reverse engineering of permanent magnets in order to find the topology from field measurements. Furthermore, a magnetic system that generate a linear external field above the magnet is presented. With a volume constraint the amount of magnetic material can be minimized without losing performance. Simulations and measurements of the printed system show a very good agreement.
\end{abstract}

\maketitle

\section*{Introduction}
Polymer bonded permanent magnets are used in a multitude of applications, ranging from sensors to various electric drive technology \cite{perm_mag_app,ehrenstein_paper,ibb}. These kinds of magnets have extremely low eddy current losses due to their low conductivity. This means, they can achieve higher efficiency compared to NdFeB sintered magnets that are usually used for radial gap motors \cite{better_motor}. For one or more dimensional sensing of position or rotation in electrical machines and mechanical components, the external field of the bias magnet is an important aspect for an accurate measurement \cite{linear}. The magnetic field of the bias magnet can be influenced by the topology and the magnetization direction. All these applications need a special designed magnetic field for their usage. Moreover, it is often desirable to reduce the amount of material required to generate the target field in order to reduce production costs, and rare-earth elements in permanent magnets \cite{reduction_ndfeb}. 

The generated external field of a given magnetic system can be designed with different methods. Examples are an inverse magnetic field computation based on a finite elements method (FEM) where the magnetization $\vec{M}$ of a defined structure is optimized \cite{inverse_flo, 3d_print_inverse}. Shape optimization improves existing designs for better performance \cite{shape_opt}. The reciprocity theorem can be employed to calculate the optimal remanent flux density of a permanent magnet system \cite{reciprocity_opt}. Parameter variation simulations can be used to find an optimal layout of predefined magnetic structures \cite{linear, opt_design}. However, all these methods needs a initial layout of the permanent magnetic system. Compared to these optimization procedures, topology optimization allows the designer of magnetic systems to find a suitable topology of the magnets from scratch. Topology optimization was initiated by mechanical and structural engineers \cite{topo_book, topo_trends}. Nowadays, it is a well established method for magnetic field problems. Possible applications include optimization of write heads of hard disks \cite{topo_harddisk}, optimization of magnetostrictive sensor applications \cite{topo_sensor}, designing of C-core actuators \cite{topo_c-core}, and optimization of rotors of brushless DC motors \cite{topo_motor}.

To proof the optimization results and to manufacture prototypes of the magnetic system, a single-unit production process is necessary. Recently, it was shown that an end-user 3D printer can be used to print polymer bonded magnets with a complex shape \cite{pub_16_1_apl, 3d_print_inverse}. 3D printing, or fused deposition modeling (FDM), is an affordable technique to manufacture models, prototypes, or end products with a minimum amount of wasted material and time. The FDM technology works by heating up wire-shaped thermoplastic filaments above the softening point. A movable extruder press the molten thermoplastic through a nozzle and builds up the object layer by layer \cite{3d-print}.

In contrast to sintered magnets, polymer bonded magnets enable the manufacturing of complex shapes and features. A disadvantage of polymer bonded magnets is their lower $(BH)_\text{max}$, what is barely half of sintered one \cite{recent_devel}. Polymer bonded magnets are composed of permanent magnetic powder with a filler content between 40~--~65~vol.\% in a polymer binder matrix. Hard magnetic ferrite (e.g. Sr, Ba), or rare-earth materials (e.g. NdFeB) are commercially available, as magnetic powder. Here, a prefabricated magnetic compound (Neofer\,\textregistered$ $ 25/60p) from Magnetfabrik Bonn GmbH is used to deduct the quality of the topology optimization results. The compound material consists of NdFeB grains with uniaxial magnetocrystalline anisotropy inside a PA11 matrix. The powder has a spherical morphology, and it is produced by employing an atomization process followed by heat treatment. The compound consists of 90~wt.\% NdFeB powder, and it has a remanence $B_r=387$~mT and a coercivity $H_{cj}=771$~kA/m \cite{pub_16_1_apl}.

In this work, a simple but accurate topology optimization simulation framework is presented. We consider the possibility of the topology optimization to reconstruct (reverse engineering) the structure of a measured magnetic field and print the result. Moreover, a special structure for a given target field is presented and physically created.

\section*{Density Method}
In this work, the density method (also known as solid isotropic microstructure with penalization (SIMP)) is used to solve the topology optimization problems. This method considers the density $\varrho$ of the material in each element, which ranges from 0 (void) to 1 (bulk). This leads to one of optimization variable per element. Intermediate densities are penalized in this approach, which means that the density of the final design should be only 0 or 1 \cite{topo_trends}. For permanent magnetic systems the magnetization of an element in the design domain $\Omega_m \subset \mathbb{R}^3$ can be formulated for the design method as:
\begin{align}
\vec{M}(\varrho)=\varrho^p\vec{M_0}
\end{align}
where $\varrho \in [0,1]$ is the density value of a FEM element, $p=1$ is the penalization parameter \cite{topo_sensor}, and $\vec{M_0}$ is the magnetization. 

The general topology optimization problem with the density method can be formulated as:
\begin{equation}
\begin{aligned}
\label{eq:min}
   &\text{Find: } \min_{\varrho} J(\varrho) \\
   &\text{subject to: } \int_{\Omega_m} \varrho(\vec{r})\text{d}\vec{r}\leq V; \\
   &0 \leq \varrho(\vec{r}) \leq 1,\,\, \vec{r} \in \Omega_m
\end{aligned}
\end{equation}
with the objective function $J$ and the maximum Volume $V$ of the design as a constraint.

A finite element method (FEM) based on the FEM library FEniCS \cite{fenics}, and the library dolfin-adjoint \cite{dolfin} for the automatic derivation of the adjoint equation of a given problem is used to solve the topology optimization. Dolfin-adjoint is a framework to solve constraint optimization problems by partial differential equations (PDE). The minimization problem is solved with the IPOPT software library for large scale nonlinear optimization problems \cite{ipopt}.

\section*{Validation of the Method}
\begin{figure}[htbp]
	\centering
	\includegraphics[width=1\linewidth]{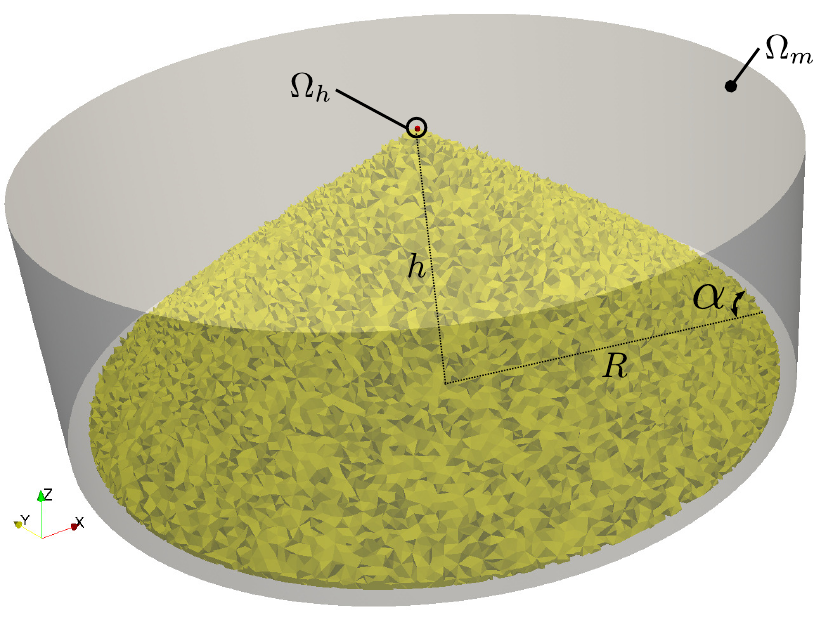}
	\caption{Maximize $B_z$ in the region $\Omega_h$ above the design volume $\Omega_m$. The resulting cone has the same angle $\alpha\approx35.2\,^\circ$ as theoretically predicted.}
	\label{fig:zylinder_max_z}
\end{figure}

To proof the effectiveness of our topology optimization framework, the magnetic field $B_z$ in a small target volume $\Omega_h$ should be maximized and compared with a theoretical consideration (Fig.~\ref{fig:zylinder_max_z}). The design domain $\Omega_m$ has the dimension $R=10$~mm, $h=7$~mm, and the target domain is a cube with the side length $a=0.1$~mm, 0.1~mm above the design domain. To maximize $B_z$ at $\Omega_h$, the objective function has the following form:
\begin{equation}
 J(\varrho)= \frac{1}{\int_{\Omega_{h}} \Arrowvert h_z(\varrho) \Arrowvert^2 \mathrm{d} \vec{r}}
\end{equation}
with the magnetic field $h_z(\varrho)$. The result of the topology optimization is a conical permanent magnet with a cone angle of $\alpha\approx35.2~^\circ$.

\begin{figure}[ht]
	\centering
	\includegraphics[width=0.9\linewidth]{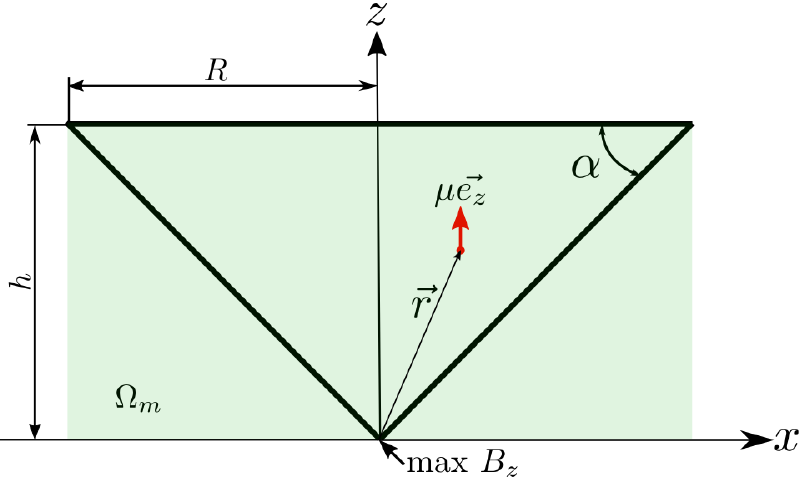}
	\caption{Sketch for the calculation of the optimal angle $\alpha$ of a conical permanent magnet.}
	\label{fig:sketch_dipol}
\end{figure}

The theoretical geometry of a magnet that creates a maximum field in a point above the magnet can be simply considered by magnetic dipoles $\vec{\mu}$ in a given design area $\Omega_m$ (Fig.~\ref{fig:sketch_dipol}). The magnetic flux density $\vec{B}$ of a magnetic dipole is \cite{jackson}:
\begin{equation}
 \vec{B}(\vec{r})=\frac{\mu_0}{4 \pi r^2} \frac{3\vec{r}(\vec{\mu}\cdot\vec{r}) - \vec{\mu}r^2}{r^3},
\end{equation}
with the vacuum permeability $\mu_0$ and the distance to the dipole $\vec{r}$. Consider $B_z$ in the $x-z$ plane, and for $\vec{\mu}\parallel \vec{e_z}$:
\begin{equation}
 B_z(\vec{r})=\frac{\mu_0 \mu\vec{e_z}}{4 \pi r^2}  \underbrace{ \left( \frac{3z^2-r^2}{r^3} \right) }_{\gamma > 0}.
\end{equation}

The field $B_z$ of a dipole is positive for $\gamma > 0$, and negative for $\gamma < 0$. Therefore, the dipoles must be positioned at all locations where this relation is fulfilled. This will lead to a maximum field component $B_z$. For the surface condition $\gamma=0$, following relation holds:
\begin{equation}
\begin{aligned}
  z&=\pm \frac{x}{\sqrt{2}}\text{, or:}\\
  \tan\alpha&=\frac{1}{\sqrt{2}}.
\end{aligned}
\end{equation}
The theoretical shape for a rotational symmetrical solid is a cone with an angle $\alpha\approx35.26^\circ$. This calculation confirms the result of the topology optimization.

\section*{Example I\\ Volume Constraint}
\begin{figure}[ht]
	\centering
	\includegraphics[width=1\linewidth]{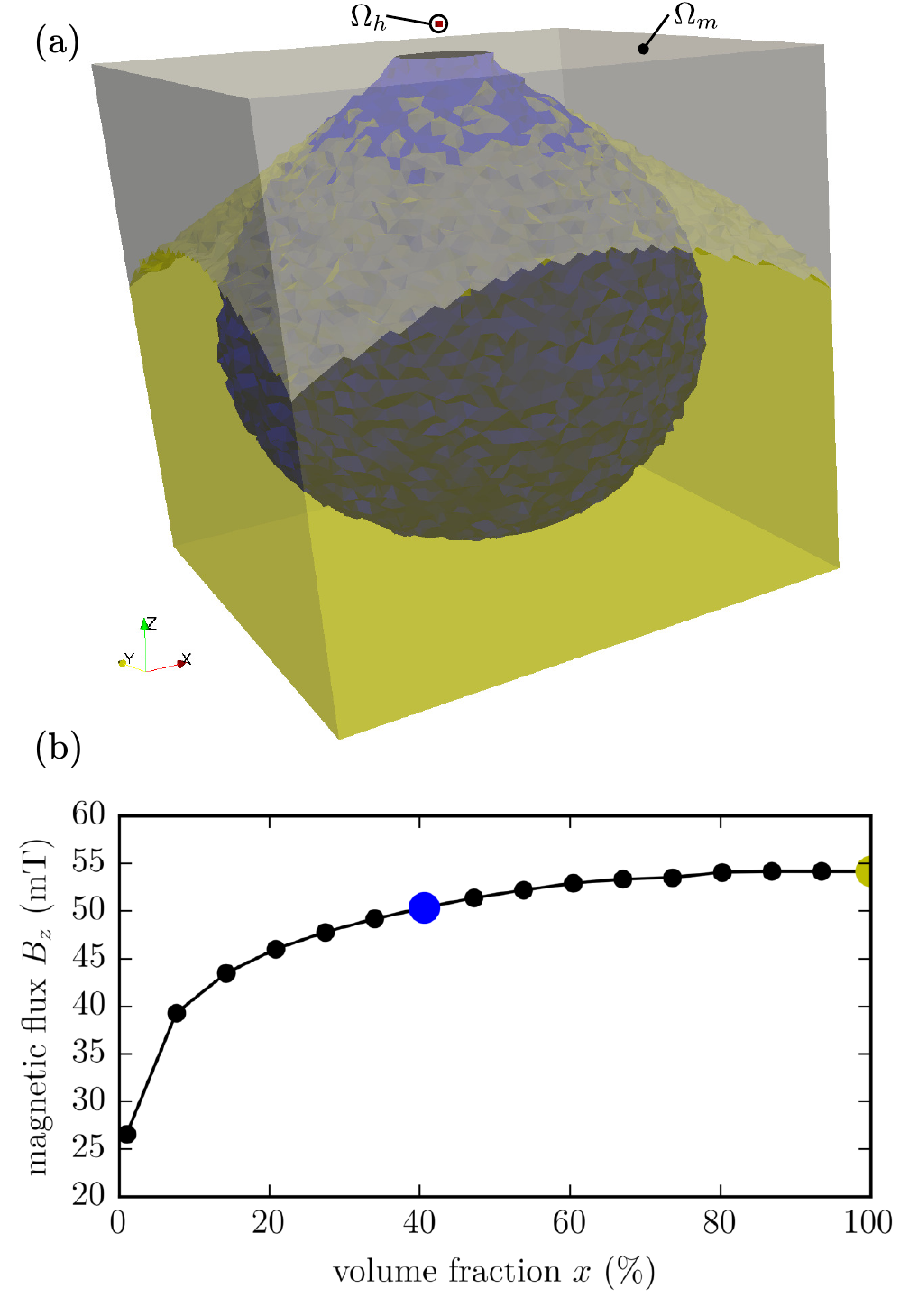}
	\caption{Maximize $B_z$ with different volume constraints of the design volume $\Omega_m$. (a) Topology of two different volume constraints ($x=40~\%$: blue, $x=100~\%$: yellow). (b) Maximum $B_z$ with different volume constraints.}
	\label{fig:max_z_vol-test}
\end{figure}

Next, the same problem as above is considered, but with a cubic design domain $\Omega_m$ with a side length $a=10$~mm, and a volume constraint. The magnetic field $B_z$ in $\Omega_h$ is simulated for different volume fractions $x=V/V_\text{max}$ in \%.  Fig.~\ref{fig:max_z_vol-test}(a) shows the geometry and the solution for two different volume fractions ($x=40~\%$: blue, $x=100~\%$: yellow). In Fig.~\ref{fig:max_z_vol-test}(b) $B_z$ is plotted in relation of the volume fraction $x\in[1,100]~\%$.

It is found that with only 40~\% of magnetic material, 94~\% of the maximum $B_z$ is attainable. These examples show that topology optimization can be used to find the topology of a permanent magnet for a given target field. Furthermore, this method can be used to reduce the amount of applied magnetic material.

\section*{Example II\\ Reverse Engineering of a Magnetic Topology}
Topology optimization is a well known method to design permanent magnets. This section shows a method for the reverse engineering of permanent magnets. This means, that the unknown geometry of a permanent magnetic system can be reconstructed from field measurements. The objective function for the minimization problem in Eq.~\ref{eq:min} can be written as:
\begin{equation}
 J(\varrho)= \int_{\Omega_{h}} \Arrowvert \vec{h_{\text{sim}}}(\varrho)-\vec{h_{\text{exp}}}\Arrowvert^2 \mathrm{d} \vec{r}
\end{equation}
where $\vec{h_{\text{sim}}}(\varrho)$ is the the simulated field, and $\vec{h_{\text{exp}}}$ is the measured external field outside the magnet.

\begin{figure*}[]
	\centering
	\includegraphics[width=1\textwidth]{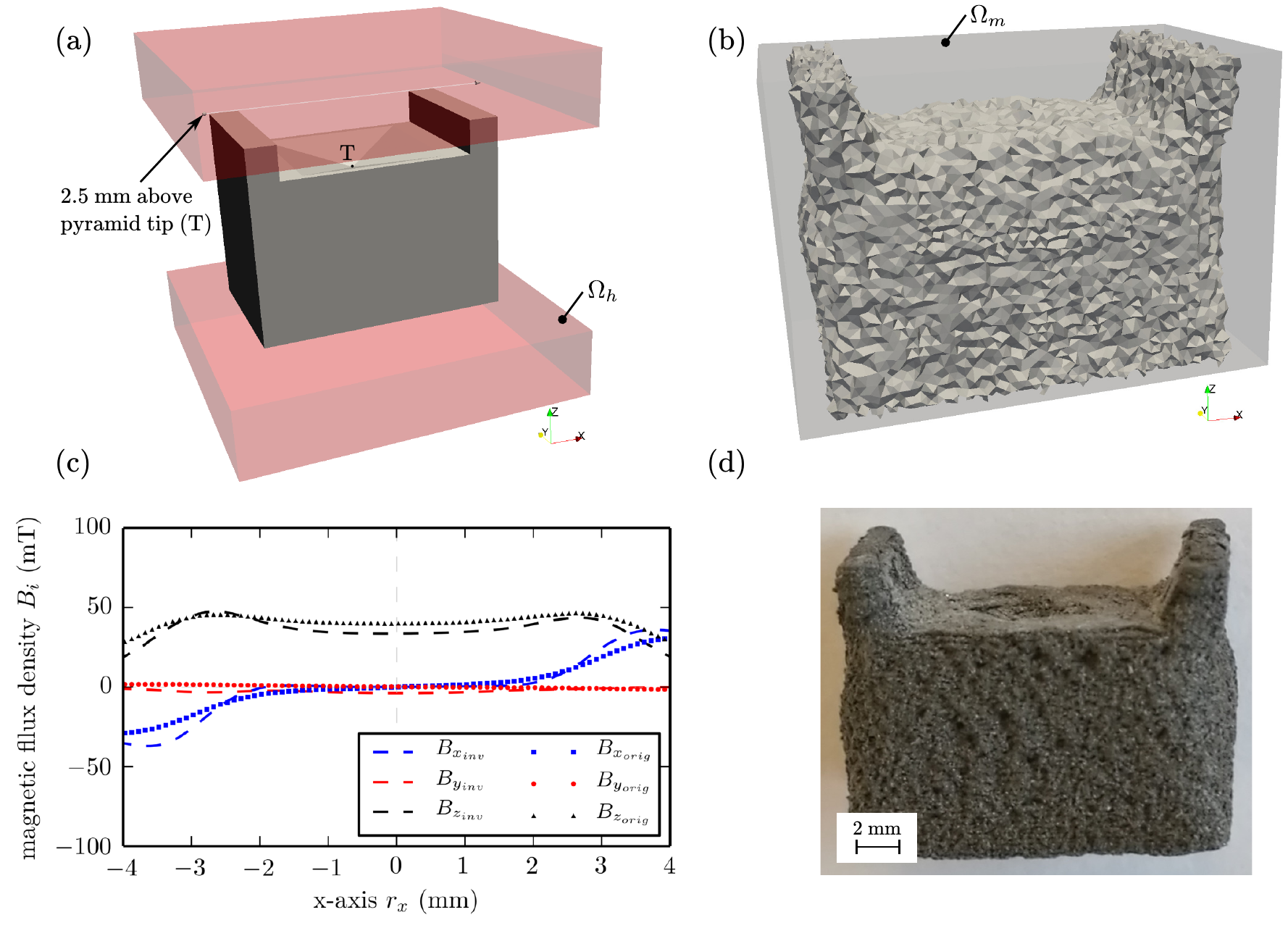}
	\caption{Reconstructed topology of a scanned magnet. (a) Geometry of a magnet that generates a specific field. The magnetic field of the field boxes $\Omega_h$ are scanned and used for the topology optimization. (b) Reconstructed magnet from measurement data. (c) Line scan of the magnetic field 2.5\,mm above the original and the reconstructed, printed magnet. (d) Picture of the printed magnet.}
	\label{fig:top_inv_bp_mag}
\end{figure*}

Recently the 3D printer process for polymer bonded magnets was benchmarked with a complex geometry that is known to minimize the components of the magnetic field $\vec{B}$ in $x$ and $y$ direction in a wide range along the $x$-axis. This is an important aspect for sensor applications \cite{pub_16_1_apl, ibb}. Here, the former printed magnet with these special characteristics is scanned with a 3D Hall sensor in a target field volume $\Omega_h$ with the size of 10$\times$10$\times$2\,mm$^3$ (L$\times$W$\times$H) 0.5~mm above and under the magnet with a resolution of 0.2~mm (Fig.~\ref{fig:top_inv_bp_mag}(a)). For the magnetic field measurements the 3D printer is upgraded to a full 3D field scanning setup \cite{pub_16_1_apl}. Fig.~\ref{fig:top_inv_bp_mag}(b) shows the design domain $\Omega_m$ with the dimension 7$\times$4.5$\times$5.5\,mm$^3$ (L$\times$W$\times$H) and the reconstructed topology of the permanent magnet. The mesh of $\Omega_m$ consists of 476225 tetrahedral elements. It is found that the reconstructed geometry has a different shape as the original one. There is no pyramid tip in the middle of the magnet and the bars on the top are not vertical. Interestingly, the volume of the reconstructed, topology optimized structure is around 5~\% lower than the volume of the original structure.

The output of the reconstruction is used to generate a model for the 3D printing process. The model is printed with the same printing parameters, and material (Neofer\,\textregistered$ $ 25/60p) from Magnetfabrik Bonn GmbH as the original one \cite{pub_16_1_apl}. The printed magnet is magnetized inside a self-built water cooled electromagnet with maximum magnetic flux density of 1.9~T in a permanent operation mode. Fig.~\ref{fig:top_inv_bp_mag}(c) displays a line scan 2.5~mm above the pyramid tip (T). Comparison between the original printed magnet and the reconstructed optimized magnets points out a good agreement. A picture of the printed, reconstructed polymer bonded magnet is shown in Fig.~\ref{fig:top_inv_bp_mag}(d).

\section*{Example III\\ Linear Position Measurement Setup}
\begin{figure}[!h]
	\centering
	\includegraphics[width=1\linewidth]{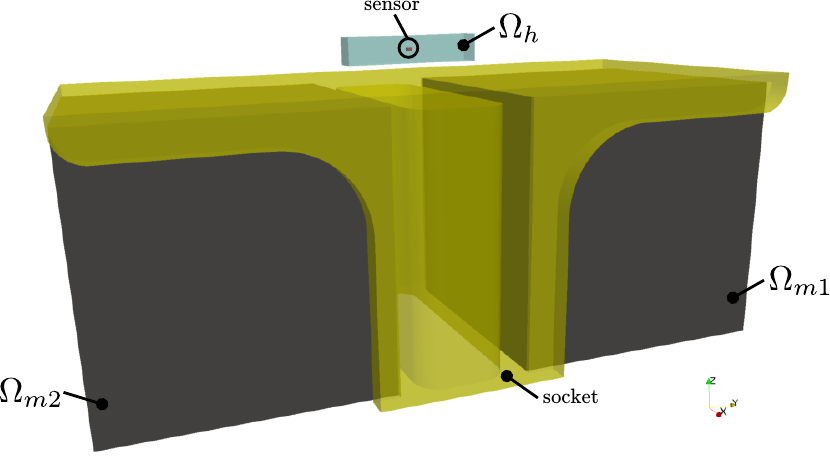}
	\caption{Setup of a magnetic configuration to linearize $B_z$ in the region $\Omega_h$.}
	\label{fig:linear_setup}
\end{figure}

\begin{figure*}[]
	\centering
	\includegraphics[width=1\textwidth]{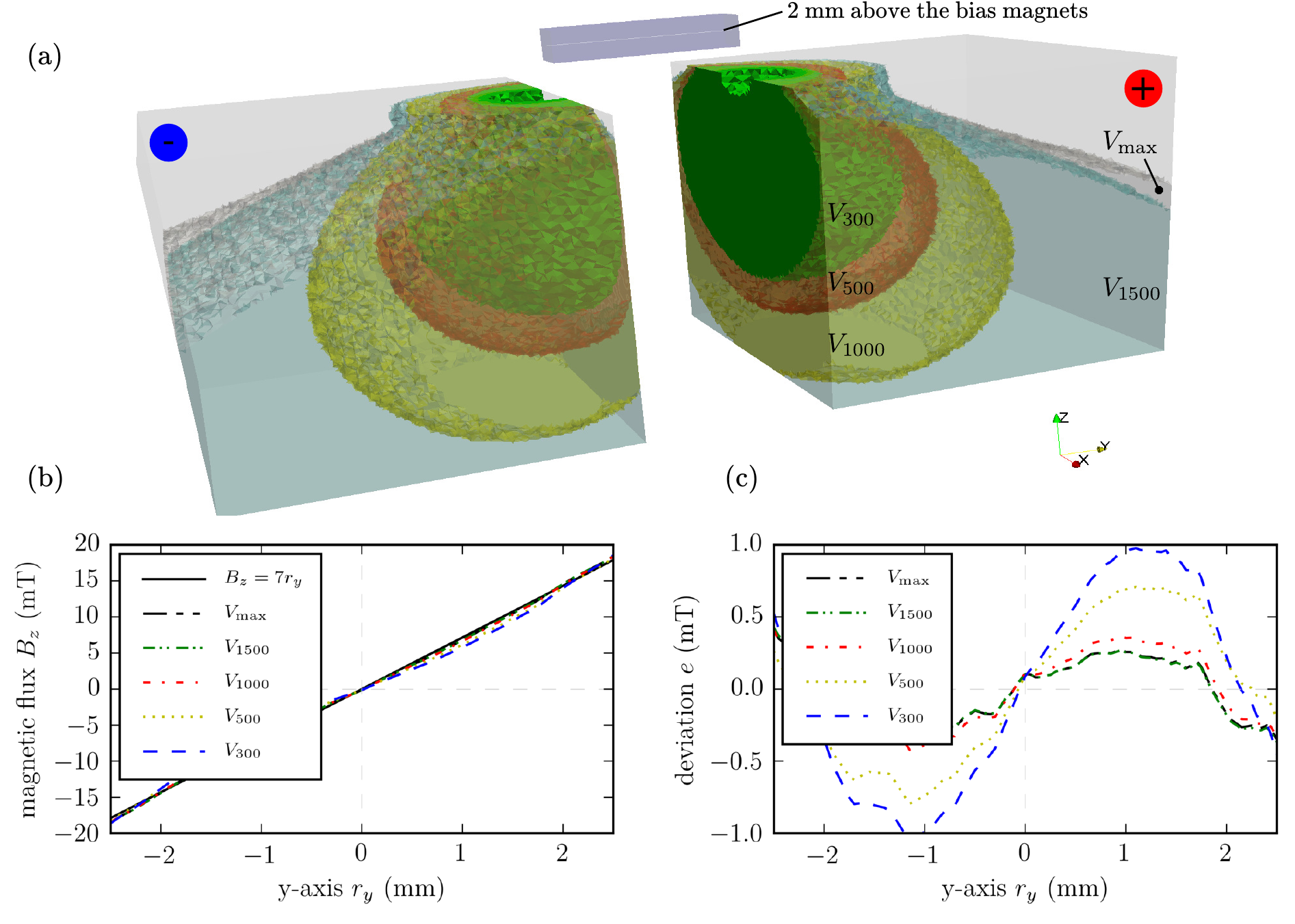}
	\caption{Topology optimized magnetic configuration with different volume constraints. (a) Picture of the topology with different volume constraints. (b) Stray field $B_z$ 2\,mm above the magnets with different volume constraints. (c) Deviation between simulation and the given linear increasing field.}
	\label{fig:linear_vol-test}
\end{figure*}
Linear position detection systems measure the position of an object in one direction. Magnetic measurement systems have some great advantages compared with other systems such as, it is a contact-free, low power, cost effective, and it is miniaturizable \cite{mag_sensor}. Currently, most magnetic measurement systems measure the field of a magnetic system and convert the sensor output signal to the position of the object. The problem is that without an optimization of the bias magnet, the position of an object as a function of the sensor output is highly non-linear. Additionally, during the manufacturing process of such measurement systems, production tolerances are unavoidable.  Parameter variation simulations for such systems already exist \cite{linear}. These kind of simulations are restricted, because they need an initial layout of the permanent magnetic system.

Here, a polymer bonded permanent magnetic system should generate a linear field in a wide range above, and along the $y$-direction of the system. Additionally, the volume should be minimized with a constraint. A model of the system is pictured in Fig.~\ref{fig:linear_setup}. It consist of a predefined not changeable socket. On both sides of the socket a cubic design domain $\Omega_{m1}$ and $\Omega_{m2}$ for the permanent magnets with a side length of $a=10$~mm are given. The gap between the two domains is 5~mm. A Hall sensor with a detection volume of 0.22$\times$0.22$\times$0.20\,mm$^3$ (L$\times$W$\times$H) is used for the determination of the position. The target field volume $\Omega_h$ has the dimension of 1$\times$5$\times$1\,mm$^3$ (L$\times$W$\times$H). Therefore, a tolerance of $\pm$0.78~mm for the sensor position is provided. The target field domain is 1.5~mm above the permanent magnetic setup.

\begin{figure*}[]
	\centering
	\includegraphics[width=1\textwidth]{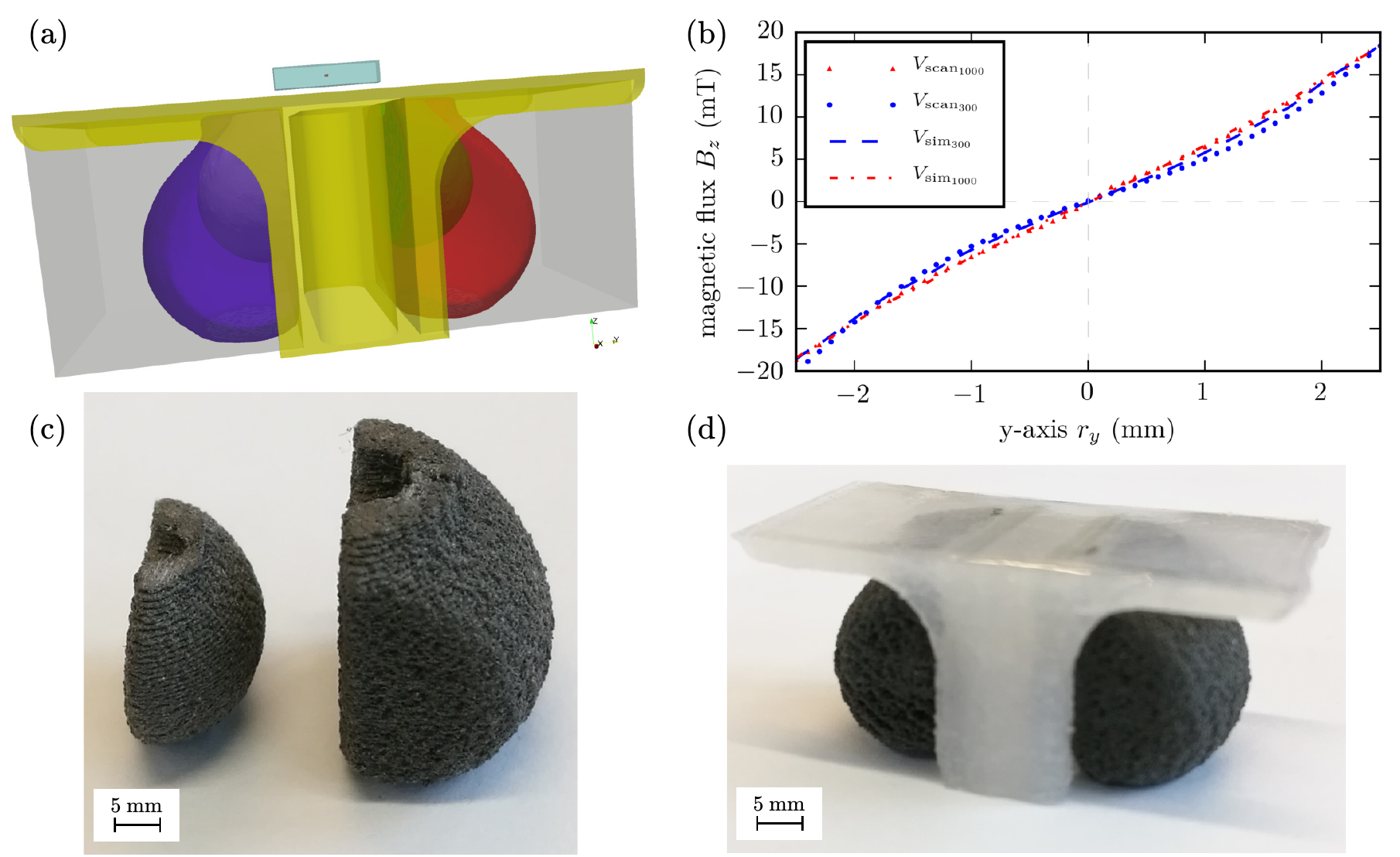}
	\caption{Comparison between simulations and measurements of the printed magnetic setup. (a) Setup for two different volume constraints ($V_\text{1000}$, $V_\text{300}$). (b) Line scan of the external field $B_z$ 2\,mm above the system compared with simulation results for both volume constraints. (c) Picture of the magnets for both constraints (right: $V_\text{1000}$, left: $V_\text{300}$). (d) Picture of the hole magnetic setup for $V_\text{1000}$.}
	\label{fig:linear_print}
\end{figure*}
The bias magnet should generate a target field $B_{z_\text{theor}}(r_z)=B_\text{max}/W r_y$ with $B_\text{max}=35$~mT. The objective function for the minimization problem in Eq.~\ref{eq:min} of this system can be formulated as:
\begin{equation}
 J(\varrho)= \int_{\Omega_{h}} \Arrowvert h_{z_\text{sim}}(\varrho)-B_{z_\text{theor}}(r_z)\Arrowvert^2 \mathrm{d} \vec{r}
\end{equation}
where $h_{z_\text{sim}}(\varrho)$ is the simulated field component in $z$ direction. 

The mesh of one design domain consist of 567888 tetrahedral elements. Simulations are performed with five different volume constraints ($V=300,~500,~1000,~1500$, and $V_\text{max}$) in mm$^3$. To create a linear magnetic field, design domain $\Omega_{m1}$ is magnetized in $z$-direction, and $\Omega_{m2}$ is magnetized in $-z$-direction. Fig.~\ref{fig:linear_vol-test}(a) shows the magnetic design with the different volume constraints. The field component $B_z$ 2~mm above the magnet for the different volume constraints is plotted in Fig.\ref{fig:linear_vol-test}(b), and the deviation of the simulated field with the theoretical field $B_{z_\text{theor}}$ is plotted in Fig.\ref{fig:linear_vol-test}(c). The solution without a volume constraint and with a maximum volume of $V=1500$~mm$^3$ is almost the same. If the maximum applicable volume decrease the deviation between the target magnetic field $B_{z_\text{theor}}$ and the simulated field increase. However, the system can be optimized in both ways. To save volume and costs or to maximize the linearity of the system. For a good compromise between applied volume and linearity of the system, a value of $V=1000$~mm$^3$ leads to a maximum deviation of  $e=0.3$~mT regarding $B_{z_\text{theor}}$ and a positioning accuracy of 0.42~\textmu m, respectively. For a low-cost system with a volume of only $V=300$~mm$^3$ the deviation increase to $e=1$~mT and a positioning accuracy of 142~\textmu m, respectively.

To verify these simulation results, the magnetic system is printed with two different volume constraints ($V=300$~mm$^3$ and $V=1000$~mm$^3$). Fig.\ref{fig:linear_print}(a) shows the system for these two volume constraints. The optimized magnets are printed with the same configuration and setup as described above. After the printing process the magnets are magnetized inside the electromagnet with a magnetic field of 1.9~T. The socket is printed with pure PA12, and the system is assembled with the two magnetized magnets. Fig.\ref{fig:linear_print}(b) displays a comparison between the printed structures and the simulated magnetic field 2~mm above the magnet. It shows a good agreement between the simulated topology optimized and the printed magnet. Fig.\ref{fig:linear_print}(c) pictures the printed magnets for both volume constraints (right: $V_\text{1000}$, left: $V_\text{300}$), and Fig.\ref{fig:linear_print}(d) from the whole magnetic system with the printed socket for $V=1000$~mm$^3$.

\section*{Conclusion}
This article presents a simple, fast, and accurate method for the topology optimization of permanent magnetic systems. As algorithm a FEM based density method is used. With a volume constraint, the amount of magnetic material can be minimized which is an important aspect in order to reduce costs and rare-earth elements in permanent magnets. The solution of the framework is confirmed with a structure that maximizes the field in the magnetization direction. The topology for such a magnet can be calculated by an analytical expression. The optimized and the theoretical structure are in a good agreement.

An additive manufacturing method of polymer bonded magnets is used to manufacture the optimized magnets. The method can be also used to recalculate and print the structure from field measurements. The magnetic field of a former 3D printed polymer bonded magnet is scanned and the topology is reconstructed. Measurements show the same magnetic field characteristic, but with a 5~\% lower volume as the original one.

Magnetically linear position detection systems have some great advantages. A problem is the high non-linearity of the sensor position and the sensor output. Here, a permanent magnetic system is optimized in order to generate a linear magnetic field in a predefined volume above the system. A volume constraint helps to save magnetic material with a minimum of performance loss. The system is printed with different volume constraints. Measurements show a perfect match with simulation results. 

\section*{Acknowledgement}
The support from CD-Laboratory AMSEN (financed by the Austrian Federal Ministry of Economy, Family and Youth, the National Foundation for Research, Technology and Development) is acknowledged. The authors would like to thank Magnetfabrik Bonn GmbH for the provision of the materials, and to Montanuniversitaet Leoben for the extrusion of the filaments. The computational results presented have been achieved using the Vienna Scientific Cluster (VSC).

\bibliographystyle{aipnum4-1}

\begin{thebibliography}{25}%
\makeatletter
\providecommand \@ifxundefined [1]{%
 \@ifx{#1\undefined}
}%
\providecommand \@ifnum [1]{%
 \ifnum #1\expandafter \@firstoftwo
 \else \expandafter \@secondoftwo
 \fi
}%
\providecommand \@ifx [1]{%
 \ifx #1\expandafter \@firstoftwo
 \else \expandafter \@secondoftwo
 \fi
}%
\providecommand \natexlab [1]{#1}%
\providecommand \enquote  [1]{``#1''}%
\providecommand \bibnamefont  [1]{#1}%
\providecommand \bibfnamefont [1]{#1}%
\providecommand \citenamefont [1]{#1}%
\providecommand \href@noop [0]{\@secondoftwo}%
\providecommand \href [0]{\begingroup \@sanitize@url \@href}%
\providecommand \@href[1]{\@@startlink{#1}\@@href}%
\providecommand \@@href[1]{\endgroup#1\@@endlink}%
\providecommand \@sanitize@url [0]{\catcode `\\12\catcode `\$12\catcode
  `\&12\catcode `\#12\catcode `\^12\catcode `\_12\catcode `\%12\relax}%
\providecommand \@@startlink[1]{}%
\providecommand \@@endlink[0]{}%
\providecommand \url  [0]{\begingroup\@sanitize@url \@url }%
\providecommand \@url [1]{\endgroup\@href {#1}{\urlprefix }}%
\providecommand \urlprefix  [0]{URL }%
\providecommand \Eprint [0]{\href }%
\providecommand \doibase [0]{http://dx.doi.org/}%
\providecommand \selectlanguage [0]{\@gobble}%
\providecommand \bibinfo  [0]{\@secondoftwo}%
\providecommand \bibfield  [0]{\@secondoftwo}%
\providecommand \translation [1]{[#1]}%
\providecommand \BibitemOpen [0]{}%
\providecommand \bibitemStop [0]{}%
\providecommand \bibitemNoStop [0]{.\EOS\space}%
\providecommand \EOS [0]{\spacefactor3000\relax}%
\providecommand \BibitemShut  [1]{\csname bibitem#1\endcsname}%
\let\auto@bib@innerbib\@empty
\bibitem [{\citenamefont {Coey}(2002)}]{perm_mag_app}%
  \BibitemOpen
  \bibfield  {author} {\bibinfo {author} {\bibfnamefont {J.}~\bibnamefont
  {Coey}},\ }\href@noop {} {\bibfield  {journal} {\bibinfo  {journal} {Journal
  of Magnetism and Magnetic Materials}\ }\textbf {\bibinfo {volume} {248}},\
  \bibinfo {pages} {441 } (\bibinfo {year} {2002})}\BibitemShut {NoStop}%
\bibitem [{\citenamefont {Eimeke}\ and\ \citenamefont
  {Ehrenstein}(2006)}]{ehrenstein_paper}%
  \BibitemOpen
  \bibfield  {author} {\bibinfo {author} {\bibfnamefont {S.}~\bibnamefont
  {Eimeke}}\ and\ \bibinfo {author} {\bibfnamefont {G.}~\bibnamefont
  {Ehrenstein}},\ }\href@noop {} {\bibfield  {journal} {\bibinfo  {journal}
  {Annual Technical Conference - ANTEC}\ }\textbf {\bibinfo {volume} {1}},\
  \bibinfo {pages} {461} (\bibinfo {year} {2006})}\BibitemShut {NoStop}%
\bibitem [{\citenamefont {Elian}\ and\ \citenamefont {Theuss}(2014)}]{ibb}%
  \BibitemOpen
  \bibfield  {author} {\bibinfo {author} {\bibfnamefont {K.}~\bibnamefont
  {Elian}}\ and\ \bibinfo {author} {\bibfnamefont {H.}~\bibnamefont {Theuss}},\
  }in\ \href@noop {} {\emph {\bibinfo {booktitle} {Electronics
  System-Integration Technology Conference (ESTC), 2014}}}\ (\bibinfo {year}
  {2014})\ pp.\ \bibinfo {pages} {1--5}\BibitemShut {NoStop}%
\bibitem [{\citenamefont {Tsunata}\ \emph {et~al.}(2016)\citenamefont
  {Tsunata}, \citenamefont {Takemoto}, \citenamefont {Ogasawara}, \citenamefont
  {Watanabe}, \citenamefont {Ueno},\ and\ \citenamefont
  {Yamada}}]{better_motor}%
  \BibitemOpen
  \bibfield  {author} {\bibinfo {author} {\bibfnamefont {R.}~\bibnamefont
  {Tsunata}}, \bibinfo {author} {\bibfnamefont {M.}~\bibnamefont {Takemoto}},
  \bibinfo {author} {\bibfnamefont {S.}~\bibnamefont {Ogasawara}}, \bibinfo
  {author} {\bibfnamefont {A.}~\bibnamefont {Watanabe}}, \bibinfo {author}
  {\bibfnamefont {T.}~\bibnamefont {Ueno}}, \ and\ \bibinfo {author}
  {\bibfnamefont {K.}~\bibnamefont {Yamada}},\ }in\ \href@noop {} {\emph
  {\bibinfo {booktitle} {2016 XXII International Conference on Electrical
  Machines (ICEM)}}}\ (\bibinfo {year} {2016})\ pp.\ \bibinfo {pages}
  {272--278}\BibitemShut {NoStop}%
\bibitem [{\citenamefont {Ortner}(2015)}]{linear}%
  \BibitemOpen
  \bibfield  {author} {\bibinfo {author} {\bibfnamefont {M.}~\bibnamefont
  {Ortner}},\ }in\ \href@noop {} {\emph {\bibinfo {booktitle} {2015 9th
  International Conference on Sensing Technology (ICST)}}}\ (\bibinfo {year}
  {2015})\ pp.\ \bibinfo {pages} {359--364}\BibitemShut {NoStop}%
\bibitem [{\citenamefont {Burkhardt}(2015)}]{reduction_ndfeb}%
  \BibitemOpen
  \bibfield  {author} {\bibinfo {author} {\bibfnamefont {C.}~\bibnamefont
  {Burkhardt}},\ }in\ \href@noop {} {\emph {\bibinfo {booktitle} {Pforzheimer
  Werkstofftag}}}\ (\bibinfo {year} {2015})\BibitemShut {NoStop}%
\bibitem [{\citenamefont {Bruckner}\ \emph {et~al.}(2017)\citenamefont
  {Bruckner}, \citenamefont {Abert}, \citenamefont {Wautischer}, \citenamefont
  {Huber}, \citenamefont {Vogler}, \citenamefont {Hinze},\ and\ \citenamefont
  {Suess}}]{inverse_flo}%
  \BibitemOpen
  \bibfield  {author} {\bibinfo {author} {\bibfnamefont {F.}~\bibnamefont
  {Bruckner}}, \bibinfo {author} {\bibfnamefont {C.}~\bibnamefont {Abert}},
  \bibinfo {author} {\bibfnamefont {G.}~\bibnamefont {Wautischer}}, \bibinfo
  {author} {\bibfnamefont {C.}~\bibnamefont {Huber}}, \bibinfo {author}
  {\bibfnamefont {C.}~\bibnamefont {Vogler}}, \bibinfo {author} {\bibfnamefont
  {M.}~\bibnamefont {Hinze}}, \ and\ \bibinfo {author} {\bibfnamefont
  {D.}~\bibnamefont {Suess}},\ }\href@noop {} {\bibfield  {journal} {\bibinfo
  {journal} {Scientific Reports}\ }\textbf {\bibinfo {volume} {7}},\ \bibinfo
  {pages} {40816} (\bibinfo {year} {2017})}\BibitemShut {NoStop}%
\bibitem [{\citenamefont {{Huber}}\ \emph {et~al.}(2017)\citenamefont
  {{Huber}}, \citenamefont {{Abert}}, \citenamefont {{Bruckner}}, \citenamefont
  {{Groenefeld}}, \citenamefont {{Schuschnigg}}, \citenamefont {{Teliban}},
  \citenamefont {{Vogler}}, \citenamefont {{Wautischer}}, \citenamefont
  {{Windl}},\ and\ \citenamefont {{Suess}}}]{3d_print_inverse}%
  \BibitemOpen
  \bibfield  {author} {\bibinfo {author} {\bibfnamefont {C.}~\bibnamefont
  {{Huber}}}, \bibinfo {author} {\bibfnamefont {C.}~\bibnamefont {{Abert}}},
  \bibinfo {author} {\bibfnamefont {F.}~\bibnamefont {{Bruckner}}}, \bibinfo
  {author} {\bibfnamefont {M.}~\bibnamefont {{Groenefeld}}}, \bibinfo {author}
  {\bibfnamefont {S.}~\bibnamefont {{Schuschnigg}}}, \bibinfo {author}
  {\bibfnamefont {I.}~\bibnamefont {{Teliban}}}, \bibinfo {author}
  {\bibfnamefont {C.}~\bibnamefont {{Vogler}}}, \bibinfo {author}
  {\bibfnamefont {G.}~\bibnamefont {{Wautischer}}}, \bibinfo {author}
  {\bibfnamefont {R.}~\bibnamefont {{Windl}}}, \ and\ \bibinfo {author}
  {\bibfnamefont {D.}~\bibnamefont {{Suess}}},\ }\href@noop {} {\bibfield
  {journal} {\bibinfo  {journal} {ArXiv e-prints}\ } (\bibinfo {year}
  {2017})},\ \Eprint {http://arxiv.org/abs/1701.07703} {arXiv:1701.07703}
  \BibitemShut {NoStop}%
\bibitem [{\citenamefont {Wang}\ and\ \citenamefont {Kang}(2000)}]{shape_opt}%
  \BibitemOpen
  \bibfield  {author} {\bibinfo {author} {\bibfnamefont {S.}~\bibnamefont
  {Wang}}\ and\ \bibinfo {author} {\bibfnamefont {J.}~\bibnamefont {Kang}},\
  }\href@noop {} {\bibfield  {journal} {\bibinfo  {journal} {IEEE Transactions
  on Magnetics}\ }\textbf {\bibinfo {volume} {36}},\ \bibinfo {pages} {1119}
  (\bibinfo {year} {2000})}\BibitemShut {NoStop}%
\bibitem [{\citenamefont {Insinga}\ \emph {et~al.}(2016)\citenamefont
  {Insinga}, \citenamefont {Bj\o{}rk}, \citenamefont {Smith},\ and\
  \citenamefont {Bahl}}]{reciprocity_opt}%
  \BibitemOpen
  \bibfield  {author} {\bibinfo {author} {\bibfnamefont {A.~R.}\ \bibnamefont
  {Insinga}}, \bibinfo {author} {\bibfnamefont {R.}~\bibnamefont {Bj\o{}rk}},
  \bibinfo {author} {\bibfnamefont {A.}~\bibnamefont {Smith}}, \ and\ \bibinfo
  {author} {\bibfnamefont {C.~R.~H.}\ \bibnamefont {Bahl}},\ }\href@noop {}
  {\bibfield  {journal} {\bibinfo  {journal} {Phys. Rev. Applied}\ }\textbf
  {\bibinfo {volume} {5}},\ \bibinfo {pages} {064014} (\bibinfo {year}
  {2016})}\BibitemShut {NoStop}%
\bibitem [{\citenamefont {Klevets}(2006)}]{opt_design}%
  \BibitemOpen
  \bibfield  {author} {\bibinfo {author} {\bibfnamefont {N.~I.}\ \bibnamefont
  {Klevets}},\ }\href@noop {} {\bibfield  {journal} {\bibinfo  {journal}
  {Journal of Magnetism and Magnetic Materials}\ }\textbf {\bibinfo {volume}
  {306}},\ \bibinfo {pages} {281 } (\bibinfo {year} {2006})}\BibitemShut
  {NoStop}%
\bibitem [{\citenamefont {Bendsoe}\ and\ \citenamefont
  {Sigmund}(2004)}]{topo_book}%
  \BibitemOpen
  \bibfield  {author} {\bibinfo {author} {\bibfnamefont {M.~P.}\ \bibnamefont
  {Bendsoe}}\ and\ \bibinfo {author} {\bibfnamefont {O.}~\bibnamefont
  {Sigmund}},\ }\href@noop {} {\emph {\bibinfo {title} {Topology
  Optimization}}}\ (\bibinfo  {publisher} {Springer},\ \bibinfo {year}
  {2004})\BibitemShut {NoStop}%
\bibitem [{\citenamefont {Campelo}, \citenamefont {Ram{\i}rez},\ and\
  \citenamefont {Igarashi}(2010)}]{topo_trends}%
  \BibitemOpen
  \bibfield  {author} {\bibinfo {author} {\bibfnamefont {F.}~\bibnamefont
  {Campelo}}, \bibinfo {author} {\bibfnamefont {J.}~\bibnamefont {Ram{\i}rez}},
  \ and\ \bibinfo {author} {\bibfnamefont {H.}~\bibnamefont {Igarashi}},\
  }\href@noop {} {\  (\bibinfo {year} {2010})}\BibitemShut {NoStop}%
\bibitem [{\citenamefont {Okamoto}, \citenamefont {Akiyama},\ and\
  \citenamefont {Takahashi}(2006)}]{topo_harddisk}%
  \BibitemOpen
  \bibfield  {author} {\bibinfo {author} {\bibfnamefont {Y.}~\bibnamefont
  {Okamoto}}, \bibinfo {author} {\bibfnamefont {K.}~\bibnamefont {Akiyama}}, \
  and\ \bibinfo {author} {\bibfnamefont {N.}~\bibnamefont {Takahashi}},\
  }\href@noop {} {\bibfield  {journal} {\bibinfo  {journal} {IEEE Transactions
  on Magnetics}\ }\textbf {\bibinfo {volume} {42}},\ \bibinfo {pages} {1087}
  (\bibinfo {year} {2006})}\BibitemShut {NoStop}%
\bibitem [{\citenamefont {Choi}\ and\ \citenamefont {Yoo}(2009)}]{topo_sensor}%
  \BibitemOpen
  \bibfield  {author} {\bibinfo {author} {\bibfnamefont {J.~S.}\ \bibnamefont
  {Choi}}\ and\ \bibinfo {author} {\bibfnamefont {J.}~\bibnamefont {Yoo}},\
  }\href@noop {} {\bibfield  {journal} {\bibinfo  {journal} {Computer Methods
  in Applied Mechanics and Engineering}\ }\textbf {\bibinfo {volume} {198}},\
  \bibinfo {pages} {2111 } (\bibinfo {year} {2009})}\BibitemShut {NoStop}%
\bibitem [{\citenamefont {Wang}\ and\ \citenamefont
  {Kang}(2002)}]{topo_c-core}%
  \BibitemOpen
  \bibfield  {author} {\bibinfo {author} {\bibfnamefont {S.}~\bibnamefont
  {Wang}}\ and\ \bibinfo {author} {\bibfnamefont {J.}~\bibnamefont {Kang}},\
  }\href@noop {} {\bibfield  {journal} {\bibinfo  {journal} {IEEE Transactions
  on Magnetics}\ }\textbf {\bibinfo {volume} {38}},\ \bibinfo {pages} {1029}
  (\bibinfo {year} {2002})}\BibitemShut {NoStop}%
\bibitem [{\citenamefont {Wang}\ \emph {et~al.}(2005)\citenamefont {Wang},
  \citenamefont {Youn}, \citenamefont {Moon},\ and\ \citenamefont
  {Kang}}]{topo_motor}%
  \BibitemOpen
  \bibfield  {author} {\bibinfo {author} {\bibfnamefont {S.}~\bibnamefont
  {Wang}}, \bibinfo {author} {\bibfnamefont {D.}~\bibnamefont {Youn}}, \bibinfo
  {author} {\bibfnamefont {H.}~\bibnamefont {Moon}}, \ and\ \bibinfo {author}
  {\bibfnamefont {J.}~\bibnamefont {Kang}},\ }\href@noop {} {\bibfield
  {journal} {\bibinfo  {journal} {IEEE Transactions on Magnetics}\ }\textbf
  {\bibinfo {volume} {41}},\ \bibinfo {pages} {1808} (\bibinfo {year}
  {2005})}\BibitemShut {NoStop}%
\bibitem [{\citenamefont {Huber}\ \emph {et~al.}(2016)\citenamefont {Huber},
  \citenamefont {Abert}, \citenamefont {Bruckner}, \citenamefont {Groenefeld},
  \citenamefont {Muthsam}, \citenamefont {Schuschnigg}, \citenamefont {Sirak},
  \citenamefont {Thanhoffer}, \citenamefont {Teliban}, \citenamefont {Vogler},
  \citenamefont {Windl},\ and\ \citenamefont {Suess}}]{pub_16_1_apl}%
  \BibitemOpen
  \bibfield  {author} {\bibinfo {author} {\bibfnamefont {C.}~\bibnamefont
  {Huber}}, \bibinfo {author} {\bibfnamefont {C.}~\bibnamefont {Abert}},
  \bibinfo {author} {\bibfnamefont {F.}~\bibnamefont {Bruckner}}, \bibinfo
  {author} {\bibfnamefont {M.}~\bibnamefont {Groenefeld}}, \bibinfo {author}
  {\bibfnamefont {O.}~\bibnamefont {Muthsam}}, \bibinfo {author} {\bibfnamefont
  {S.}~\bibnamefont {Schuschnigg}}, \bibinfo {author} {\bibfnamefont
  {K.}~\bibnamefont {Sirak}}, \bibinfo {author} {\bibfnamefont
  {R.}~\bibnamefont {Thanhoffer}}, \bibinfo {author} {\bibfnamefont
  {I.}~\bibnamefont {Teliban}}, \bibinfo {author} {\bibfnamefont
  {C.}~\bibnamefont {Vogler}}, \bibinfo {author} {\bibfnamefont
  {R.}~\bibnamefont {Windl}}, \ and\ \bibinfo {author} {\bibfnamefont
  {D.}~\bibnamefont {Suess}},\ }\href@noop {} {\bibfield  {journal} {\bibinfo
  {journal} {Applied Physics Letters}\ }\textbf {\bibinfo {volume} {109}},\
  \bibinfo {pages} {162401} (\bibinfo {year} {2016})}\BibitemShut {NoStop}%
\bibitem [{\citenamefont {Guo}\ and\ \citenamefont {Leu}(2013)}]{3d-print}%
  \BibitemOpen
  \bibfield  {author} {\bibinfo {author} {\bibfnamefont {N.}~\bibnamefont
  {Guo}}\ and\ \bibinfo {author} {\bibfnamefont {M.~C.}\ \bibnamefont {Leu}},\
  }\href@noop {} {\bibfield  {journal} {\bibinfo  {journal} {Frontiers of
  Mechanical Engineering}\ }\textbf {\bibinfo {volume} {8}},\ \bibinfo {pages}
  {215} (\bibinfo {year} {2013})}\BibitemShut {NoStop}%
\bibitem [{\citenamefont {Ma}\ \emph {et~al.}(2002)\citenamefont {Ma},
  \citenamefont {Herchenroeder}, \citenamefont {Smith}, \citenamefont {Suda},
  \citenamefont {Brown},\ and\ \citenamefont {Chen}}]{recent_devel}%
  \BibitemOpen
  \bibfield  {author} {\bibinfo {author} {\bibfnamefont {B.}~\bibnamefont
  {Ma}}, \bibinfo {author} {\bibfnamefont {J.}~\bibnamefont {Herchenroeder}},
  \bibinfo {author} {\bibfnamefont {B.}~\bibnamefont {Smith}}, \bibinfo
  {author} {\bibfnamefont {M.}~\bibnamefont {Suda}}, \bibinfo {author}
  {\bibfnamefont {D.}~\bibnamefont {Brown}}, \ and\ \bibinfo {author}
  {\bibfnamefont {Z.}~\bibnamefont {Chen}},\ }\href@noop {} {\bibfield
  {journal} {\bibinfo  {journal} {Journal of Magnetism and Magnetic Materials}\
  }\textbf {\bibinfo {volume} {239}},\ \bibinfo {pages} {418 } (\bibinfo {year}
  {2002})}\BibitemShut {NoStop}%
\bibitem [{\citenamefont {Logg}(2013)}]{fenics}%
  \BibitemOpen
  \bibfield  {author} {\bibinfo {author} {\bibfnamefont {A.}~\bibnamefont
  {Logg}},\ }\href@noop {} {\emph {\bibinfo {title} {Automated Solution of
  Differential Equations by the Finite Element Method (Lecture Notes in
  Computational Science and Engineering)}}}\ (\bibinfo  {publisher}
  {Springer},\ \bibinfo {year} {2013})\BibitemShut {NoStop}%
\bibitem [{\citenamefont {{Funke}}\ and\ \citenamefont
  {{Farrell}}(2013)}]{dolfin}%
  \BibitemOpen
  \bibfield  {author} {\bibinfo {author} {\bibfnamefont {S.~W.}\ \bibnamefont
  {{Funke}}}\ and\ \bibinfo {author} {\bibfnamefont {P.~E.}\ \bibnamefont
  {{Farrell}}},\ }\href@noop {} {\bibfield  {journal} {\bibinfo  {journal}
  {ArXiv e-prints}\ } (\bibinfo {year} {2013})},\ \Eprint
  {http://arxiv.org/abs/1302.3894} {arXiv:1302.3894} \BibitemShut {NoStop}%
\bibitem [{\citenamefont {W{\"a}chter}\ and\ \citenamefont
  {Biegler}(2006)}]{ipopt}%
  \BibitemOpen
  \bibfield  {author} {\bibinfo {author} {\bibfnamefont {A.}~\bibnamefont
  {W{\"a}chter}}\ and\ \bibinfo {author} {\bibfnamefont {L.~T.}\ \bibnamefont
  {Biegler}},\ }\href@noop {} {\bibfield  {journal} {\bibinfo  {journal}
  {Mathematical Programming}\ }\textbf {\bibinfo {volume} {106}},\ \bibinfo
  {pages} {25} (\bibinfo {year} {2006})}\BibitemShut {NoStop}%
\bibitem [{\citenamefont {Jackson}(1999)}]{jackson}%
  \BibitemOpen
  \bibfield  {author} {\bibinfo {author} {\bibfnamefont {J.}~\bibnamefont
  {Jackson}},\ }\href@noop {} {\emph {\bibinfo {title} {Classical
  electrodynamics}}}\ (\bibinfo  {publisher} {Wiley},\ \bibinfo {address} {New
  York},\ \bibinfo {year} {1999})\BibitemShut {NoStop}%
\bibitem [{\citenamefont {Treutler}(2001)}]{mag_sensor}%
  \BibitemOpen
  \bibfield  {author} {\bibinfo {author} {\bibfnamefont {C.}~\bibnamefont
  {Treutler}},\ }\href@noop {} {\bibfield  {journal} {\bibinfo  {journal}
  {Sensors and Actuators A: Physical}\ }\textbf {\bibinfo {volume} {91}},\
  \bibinfo {pages} {2 } (\bibinfo {year} {2001})},\ \bibinfo {note} {third
  European Conference on Magnetic Sensors \&amp; Actuators.}\BibitemShut
  {Stop}%
\end{thebibliography}

%

\end{document}